\documentclass[acmsmall]{acmart}
\AtBeginDocument{%
  }

\setcopyright{acmlicensed}
\copyrightyear{2018}
\acmYear{2018}
\acmDOI{XXXXXXX.XXXXXXX}

\acmJournal{JACM}
\acmVolume{37}
\acmNumber{4}
\acmArticle{111}
\acmMonth{8}

\usepackage{algorithmicx}
\usepackage{algpseudocode}
\usepackage{algorithm}
\usepackage{booktabs}

\begin{document}

\title{One-shot Face Sketch Synthesis in the Wild via Generative Diffusion Prior and Instruction Tuning}


\author{Han Wu}
\orcid{0009-0005-8371-9997}
\email{2112303125@mail2.gdut.edu.cn}
\affiliation{%
  \institution{Guangdong University of Technology}
  \city{Guangzhou}
  \state{Guangdong}
  \country{China}}

\author{Junyao Li}
\orcid{0009-0002-4541-5114}
\email{2112403037@mail2.gdut.edu.cn}
\affiliation{%
  \institution{Guangdong University of Technology}
  \city{Guangzhou}
  \state{Guangdong}
  \country{China}}

\author{Kangbo Zhao}
\orcid{0009-0009-7479-7757}
\email{zhaokangbo329@gmail.com}
\affiliation{%
  \institution{Guangdong University of Technology}
  \city{Guangzhou}
  \state{Guangdong}
  \country{China}}
  
\author{Sen Zhang}
\orcid{0000-0002-8065-5095}
\email{senzhang.thu10@gmail.com}
\affiliation{%
  \institution{TikTok, ByteDance}
  \city{Sydney}
  \country{Australia}}
  
\author{Yukai Shi}
\authornote{Corresponding author}
\orcid{0000-0002-9413-6528}
\email{ykshi@gdut.edu.cn}
\affiliation{%
  \institution{Guangdong University of Technology}
  \city{Guangzhou}
  \state{Guangdong}
  \country{China}}
  
\author{Liang Lin}
\orcid{0000-0001-5704-4168}
\email{linliang@ieee.org}
\affiliation{%
  \institution{Sun Yat-sen University}
  \city{Guangzhou}
  \state{Guangdong}
  \country{China}}


\begin{abstract}
  Face sketch synthesis is a technique aimed at converting face photos into sketches. Existing face sketch synthesis research mainly relies on training with numerous photo-sketch sample pairs from existing datasets. However, these large-scale discriminative learning methods will have to face problems such as data scarcity and high human labor costs. Once the training data becomes scarce, their generative performance significantly degrades. In this paper, we propose a one-shot face sketch synthesis method based on diffusion models. We optimize text instructions on a diffusion model using face photo-sketch image pairs. Then, the instructions derived through gradient-based optimization are used for inference. To simulate real-world scenarios more accurately and evaluate method effectiveness  more comprehensively, we introduce a new benchmark named One-shot Face Sketch Dataset (OS-Sketch). The benchmark consists of 400 pairs of face photo-sketch images, including sketches with different styles and photos with different backgrounds, ages, sexes, expressions, illumination, etc. For a solid out-of-distribution evaluation, we select only one pair of images for training at each time, with the rest used for inference. Extensive experiments demonstrate that the proposed method can convert various photos into realistic and highly consistent sketches in a one-shot context. Compared to other methods, our approach offers greater convenience and broader applicability. The dataset will be available at: \href{https://github.com/HanWu3125/OS-Sketch}{https://github.com/HanWu3125/OS-Sketch}
\end{abstract}

\ccsdesc[500]{Human-centered computing~Social media}


\keywords{Human Facial Sketch, One-Shot, Diffusion
Model, Out-of-Distribution}

\received{20 February 2007}
\received[revised]{12 March 2009}
\received[accepted]{5 June 2009}

\maketitle

\section{Introduction}

\begin{figure}[t]
    \centering
    \includegraphics[width=\textwidth,height = 3.5cm]{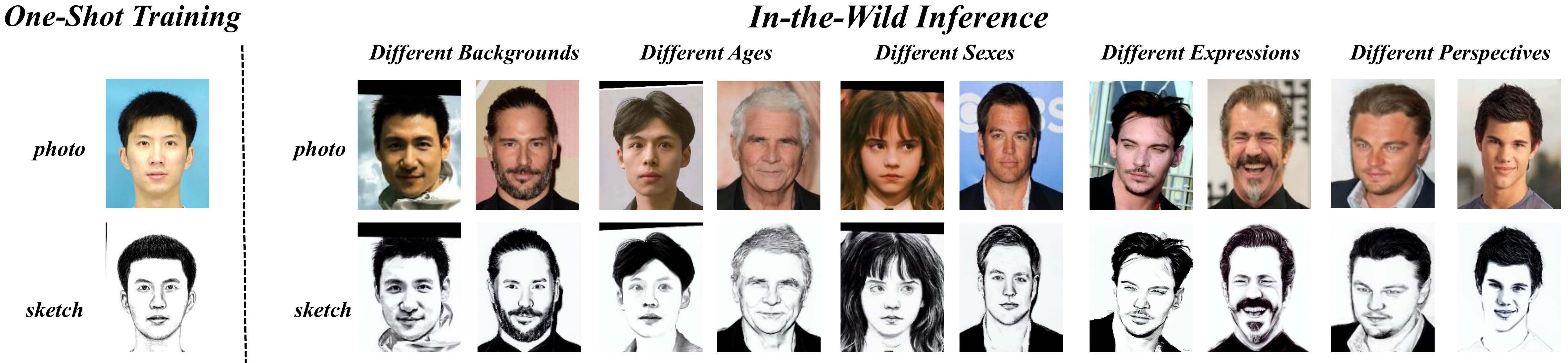}
    \caption{An illustration of one-shot sketch synthesis. We call for a solid face sketch model is able to generate realistic sketches for face photos with different backgrounds, ages, sexes, expressions, perspectives, etc. Thus, in our model, only one pair of sketch samples is used for training, while a large number of diverse images are utilized for inference. }
    \label{fig1}
\end{figure}

Face sketch synthesis is a technique aimed at converting face photos into sketches, which holds significant potential for application in digital multimedia and security field~\cite{fss5,tmm1,tomm2}. As shown in Fig.~\ref{fig1}, a variety of colorful photos are converted into realistic sketches. Existing face sketch synthesis researches~\cite{panet,gan4,gan9,fsgan,hida} already have remarkable achievements. However, they mainly rely on training with numerous photo-sketch sample pairs from existing datasets. And these large-scale discriminative learning methods will have to face problems: data scarcity and high human labor costs.

\begin{figure}[t]
    \centering
    \includegraphics[width=0.95\textwidth]{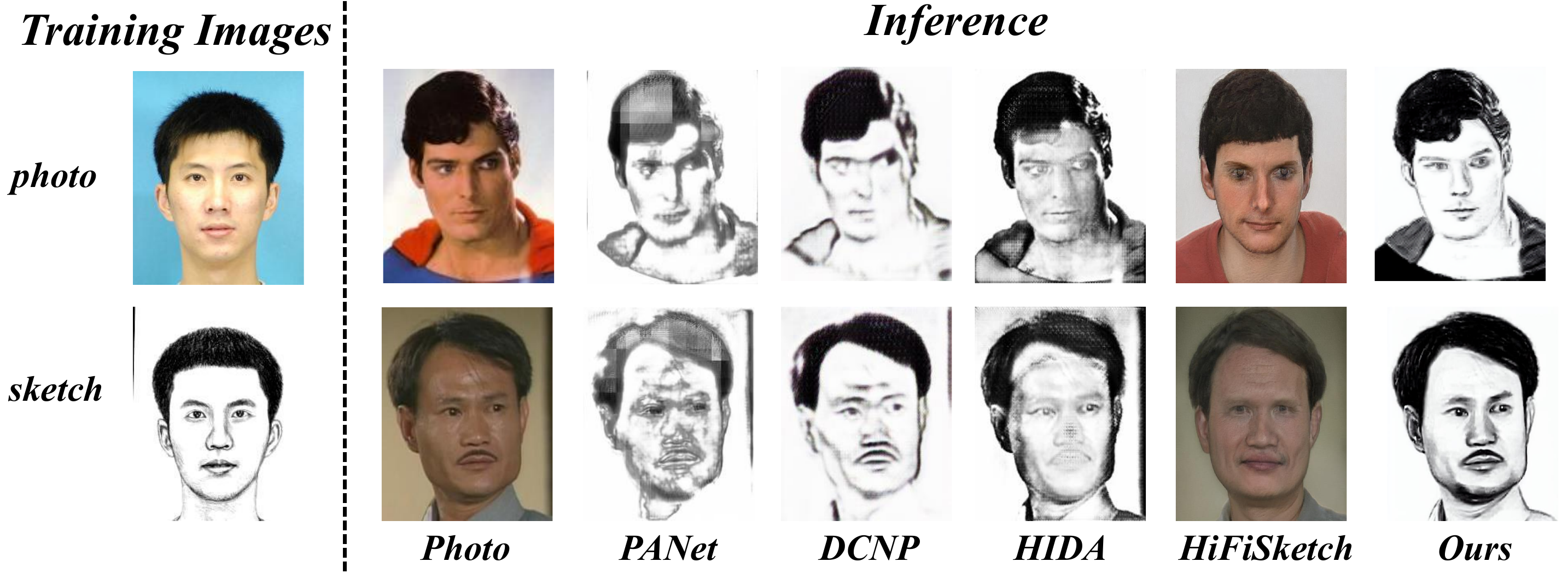}
    \caption[something short]{A demonstration of Out-of-Distribution (OOD) challenge. When trained with only a single photo-sketch pair, existing methods exhibit severe performance degradation and even fail completely when confronted with out-of-distribution(OOD) photos. In contrast, our approach can convert various out-of-distribution (OOD) photos into sketches and maintain extremely high consistency with the photos. }
    \label{fig2}
\end{figure}

 As shown in Fig.~\ref{fig2}, existing methods exhibit severe performance degradation when trained on just one photo-sketch pair. They even fail completely when confronted with out-of-distribution(OOD) photos. While traditional methods can achieve good generative capability as long as sufficient training data is available, the generalization ability often be violated when the training data is insufficient. In reality, the number of artists is limited. And the manual drawing of face sketches requires a substantial amount of time and effort. Consequently, hand-drawn sketches are costly and scarce. The limitation also results in the current situation where the number of existing datasets is limited. Nowadays, even the largest manually drawn face sketch dataset~\cite{fsgan} contains only thousands of face sketch image pairs. How to balance the high manual sketching costs with the contradiction that traditional methods rely on massive data for training remains an unresolved issue.

In recent years, diffusion models~\cite{diff1(6),ldm,glide,diff38,diffclip} have become well-established tools for generating images. Especially for text-to-image diffusion models, which are capable of producing remarkable images when provided with an image and an editing instruction. By utilizing a text-to-image diffusion model for face sketch synthesis, it is possible to generate sketches without any training, which results in significant savings in training costs. Nevertheless, it is hard to achieve because relying on text descriptions solely is fragile. Specific textures and styles of sketches and the dispersed colour representation of different areas are difficult to describe through words. Recently, image editing via visual prompting has begun to be explored~\cite{visii,vis1,vis2}. Visual Instruction Inversion (VISII)~\cite{visii} introduces a robust framework for image editing based on visual prompting. By using images as prompts instead of solely text descriptions, image editing can be more intuitive and precise.  Nonetheless, it mainly addresses basic style transfer applications for a general audience and does not explore the specific synthesis of face sketches. 

In this paper, we introduce a new framework for face sketch synthesis called One-shot Face Sketch Synthesis. Inspired by image editing via visual prompting~\cite{visii}, we optimize text instruction on a diffusion model using only a single face photo-sketch image pairs. Then, the instruction derived through gradient-based optimization is used for inference. We verify the effectiveness of our method by processing various photos in a one-shot context. As shown in Fig.~\ref{fig1}, we can generate realistic sketches with only a pair of images for training. Our method can handle photos in-the-wild with different backgrounds, ages, 
sexes, expressions, and perspectives. Our method can inherit the style of the training sketches while maintaining high consistency with the photos. At the same time, it significantly reduces energy costs. Compared to other methods, our approach offers greater convenience and broader applicability.

Moreover, we find that the existing dataset is limited by the variety of sketch styles. For example, existing datasets CUHK Face Sketch (CUFS)~\cite{cufs} dataset, CUHK Face Sketch FERET (CUFSF)~\cite{cufsf} dataset and WildSketch~\cite{panet} dataset only have a single sketch style. FS2K~\cite{fsgan} dataset is the largest available hand-drawn face photo-sketch dataset currently, but it only has 3 sketch styles. However, in real-world scenarios, sketch styles have a greater variety of diversity. There are many exquisite sketches available on the internet. Besides, we believe using a dataset combining photographic features from diverse datasets would enable more comprehensive evaluation in one-shot face sketch synthesis research. In this paper, to more accurately simulate real-world scenarios and validate the effectiveness of our method, we introduce a new benchmark named One-shot Face Sketch Dataset (OS-Sketch). First we selected 40 sketches drawn by amateurs or bloggers along with corresponding photos from the internet. Then we select 100 photo-sketch pairs from the simple color backgrounds face sketch dataset~\cite{cufs}, 110 pairs from the dark lighting backgrounds dataset~\cite{cufsf} and 150 pairs from the in-the-wild dataset~\cite{panet}. Our dataset possesses the following characteristics. First, our dataset comprises a diverse style of sketches. It not only includes sketches previously used in research but also incorporates out-of-distribution (OOD) style sketches selected from the internet. Our dataset is more complex than ever before and closer to real-world scenarios. Additionally, sketches selected from the internet are also available for researchers to conduct experimental studies. Second, the dataset not only contains traditional research photos with a single background but also includes numerous in-the-wild photos. It includes photos with different backgrounds, ages, sexes, expressions, perspectives and illumination. By exposing the model to a variety of photos in a one-shot context, its performance can be intuitively demonstrated.

To this end, the contributions of this paper can be summarized as follows:
\begin{itemize}

\item[$\bullet$]We present a new framework for face sketch synthesis. Specifically, only a pair of samples is required to ensure the achievement of face sketch synthesis in-the-wild under real-world conditions. 

\item[$\bullet$]We introduce a more challenging benchmark One-shot Face Sketch Dataset (OS-Sketch), which contains a total of 400 face photo-sketch image pairs, encompassing sketches with various styles and photos with different backgrounds, sexes, expressions, ages, illumination, perspectives, etc.


\item[$\bullet$]Our work will significantly facilitate the development of face sketch tasks in real-world scenarios as former works focus on datasets with limited variation. Extensive experiments demonstrate that the proposed method can convert various photos into realistic and highly consistent sketches in a one-shot context.
\end{itemize}

\section{RELATED WORK}
\label{related work}
\subsection{Face Photo-Sketch Synthesis}
Face photo-sketch synthesis is an important task in the field of computer vision~\cite{fss1,cufs,fss2}, which aims to convert face photos into sketches.  Traditional methods are mainly data-driven~\cite{fss2}, relying on searching for similar blocks and linear combination weight computation. Sample-based approaches~\cite{fss3,fss4,fss5} have proven to be quite effective. And regression-based approaches~\cite{fss6,fss7} produce synthesized images with more details. Nevertheless, traditional methods tend to be computationally intensive since they need to search through the entire training set to find the nearest neighbours. 

In recent years, with the development of deep learning, many researchers have applied deep learning to face photo-sketch synthesis task~\cite{fss8,fss9,fss10}. For example, Zhang et al.~\cite{fss11} use an end-to-end fully convolutional network for the task of generating sketches from photos. Zhang et al.~\cite{fss12} further proposed a content-adaptive sketch generation method, which is achieved by using a branched fully convolutional network (BFCN) to learn multi-level and multi-scale decomposition representations of facial images. Jiao et al.~\cite{fss13} proposed a lightweight four-layer CNN method for generating face sketches, which can retain most of the details. Chen et al.~\cite{fss14} proposed a pyramid column feature based on CNN, which enriches the sketch textures and shadings. With the advancement of generative adversarial networks (GANs)~\cite{gan1,gan2} and their remarkable success in diverse image generation tasks~\cite{gan3,tomm5,tomm6}, researchers have begun applying GANs to face photo-sketch synthesis. For example, Philip et al.~\cite{gan4} attempted to apply conditional generative adversarial networks (cGANs)~\cite{gan6} to the synthesis of face sketches. Wang et al.~\cite{gan5} further refined their approach by combining the original cGANs~\cite{gan6} with a post-processing method called back-projection to produce more detailed sketches. Yi et al.~\cite{gan7} propose a composite GAN architecture called APDrawingGAN++, which consists of local networks to learn effective representations for specific facial features and a global network to capture the overall content. Yu et al.~\cite{gan8} introduced a new framework named Composition-Aided Generative Adversarial Network (CA-GAN), which leverages facial composition information to enhance the synthesis process, ensuring the authenticity of structure and consistency of texture in the generated facial sketches. Peng et al.~\cite{gan9} proposed a new cross-domain face photo-sketch synthesis framework named HiFiSketch. It learns to adjust the generator weights to achieve high-fidelity synthesis and manipulation, allowing for the translation of images between the photo domain and the sketch domain while enabling modifications according to the text input.  
 
Despite achieving certain results, previous methods require extensive datasets for training, and obtain weak generalization capabilities on out-of-distribution cases. In this paper, we present a one-shot face sketch synthesis method with generative diffusion prior, which fully addresses the generalization capabilities toward out-of-distribution cases. 

\subsection{Diffusion Models for Image Editing}
Image editing is a technique that enables users to manipulate images based on their own expressions~\cite{zero,tmm4}. Nowadays, diffusion models have garnered significant accomplishments in image generation~\cite{diff1(6),ldm,tomm1,tomm3,tomm4,tomm7} and have been extended to image editing~\cite{diff1,insp2p,diff2,diff3,diff4}. SDEdit~\cite{sdedit} applies a pre-trained model to add noise to the input image and denoise it with a new user editing guide. Later, GLIDE~\cite{glide} and Stable Diffusion models~\cite{ldm} incorporate this work into text-to-image generation, enabling effective text-based image editing. DiffusionCLIP~\cite{diffclip} combines diffusion models with Contrastive Language-Image Pre-training (CLIP)~\cite{clip} to perform zero-shot image manipulation and their method of image manipulation successfully performs both in the trained and unseen domain. Imagic~\cite{imagic} tried to leverage pre-trained diffusion models for image editing, but it necessitated fine-tuning of the model. In contrast, prompt-to-prompt~\cite{diff2} eliminates the need for training or fine-tuning of the model. It constitutes an intuitive image editing interface through editing only the textual prompt. Pix2pix-zero~\cite{zero} introduces a diffusion-based image-to-image translation method that eliminates the need for training or prompts. Users simply specify the editing direction from the source to the target domain (such as cat → dog) without the need for manually crafting text prompts for the input image. Visual Instruction Inversion (VISII)~\cite{visii} introduces a visual prompt-based approach to image editing, allowing users to replace hard-to-describe editing tasks with corresponding images. 

Due to the powerful generative capabilities of diffusion models, in this paper we adopt a strategy that combines diffusion models with face sketch synthesis. Compared to previous methods, our method greatly reduces training costs while ensuring generation quality.

\section{Methodology}
\begin{figure}[!t]
    \centering
    \includegraphics[width=\textwidth,height=7cm]{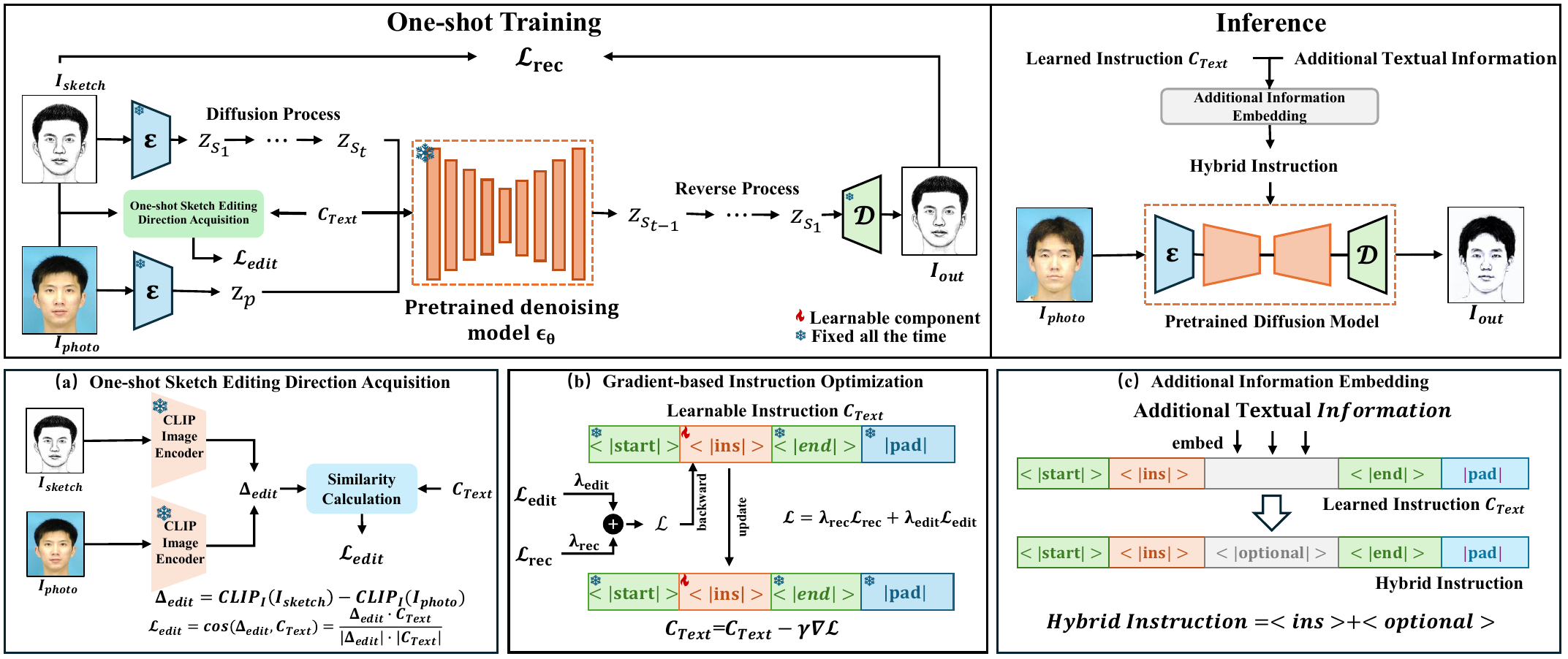}
    \caption{Our framework jointly optimizes the text instruction $C _{Text}$ through the image reconstruction strategy of a diffusion model for one-shot sketch synthesis. Here, $\mathcal{E}$ denotes the encoder, and $\mathcal{D}$ denotes the decoder. (a) We encode a pair of sketch samples $\lbrace I _{photo}, I_{sketch}, \rbrace$ into the embedding space using CLIP~\cite{clip}. Then, we calculate the CLIP embedding distance $\bigtriangleup _{edit}$ between $I _{photo}$ and $I _{sketch}$, and utilize $\bigtriangleup _{edit}$ to optimize the text instruction $C _{Text}$. In this way, $C _{Text}$ is able to represent the editing direction from a photo to a sketch. We choose cosine similarity for the optimization process. (b) We perform gradient-based optimization on the learnable Instruction $C _{Text}$. We only optimize a part of the instruction, marked as $<ins>$. We use two losses $\mathcal{L} _{rec}$ and $\mathcal{L} _{edit}$ to jointly optimize $C _{Text}$. (c) During inference, we can embed additional textual information into learned instruction $C _{Text}$. The embedded part is marked as $<optional>$. The final hybrid instruction is formulated by integrating the optimized $<ins>$ with the supplementary $<optional>$ provided more historical context. }
    \label{fig3}
\end{figure}

In this section, we mainly introduce our method. First, we introduce a brief background on text-to-image diffusion models. Then, we present the training framework for One-shot Face Sketch Synthesis. The framework of our method is shown in Fig.~\ref{fig3}.

\subsection{Preliminaries}
Diffusion models learn to generate data samples through a denoising sequence that estimates the score of the data distribution. In the forward process, Gaussian noise $\epsilon$ is gradually added to the input image $x$ in T steps, producing a series of noisy samples $x _{1}$, ..., $x _{T}$. In the backward process, $x _{T}$ will be restored to $x$ through a denoising neural network. 

Initially, diffusion models were operated in pixel space~\cite{diff1(6)}. To enable diffusion models training on limited computational resources while retaining their quality and flexibility, Latent Diffusion Model (LDM)~\cite{ldm} proposes running the diffusion process in latent space. Latent Diffusion Model (LDM)~\cite{ldm} introduces an autoencoder model for image reconstruction. The encoder $\mathcal{E}$ encodes the input image into a 2D latent space $z$ to obtain the latent image $z _{x}$. And the decoder $\mathcal{D}$  is used for decoding. To enable the diffusion model to generate more accurately, LDM~\cite{ldm}  also introduces a domain-specific encoder $\tau _{\theta}$, which projects text prompts into an intermediate representation $C _{Text}$. Then, $C _{Text}$ is inserted into the layers of the denoising network through cross-attention as an instruction of the model. Thus, the diffusion model is allowed to take text prompts as conditional input. The objective function is:

\begin{equation}
\label{eq1}
\mathcal{L}=\mathbb{E}_{\mathcal{E}(x), C_{Text}, \epsilon \sim \mathcal{N}(0,1), t}\left\|\epsilon-\epsilon_{\theta}\left(z_{x_{t}}, t, C_{Text}\right)\right\|_{2}
\end{equation}
where t denotes each time step.

However, text instructions $C _{Text}$ are hard to align with the desired edits precisely. This limitation causes diffusion-generated images to struggle to fully adhere to the pixel-level details of the input images. To address this challenge, InstructPix2Pix~\cite{insp2p} proposes incorporating the input image into the denoising network. Specifically, the input image $x$ is encoded into a conditional image $\mathcal{E}(x)$ and concatenated with the latent image $z _{x _{t}}$. As a result, the objective function Eq.~\ref{eq1} is modified to:
\begin{equation}
\label{eq2}
\mathcal{L}=\mathbb{E}_{\mathcal{E}(x), C_{Text},\mathcal{E}(x), \epsilon \sim \mathcal{N}(0,1), t}\left\|\epsilon-\epsilon_{\theta}\left(z_{x _{t}}, t, C_{Text},\mathcal{E}(x)\right)\right\|_{2}
\end{equation}

\subsection{Framework}
\begin{figure*}[!t]
\centering
\includegraphics[width=0.99\linewidth]{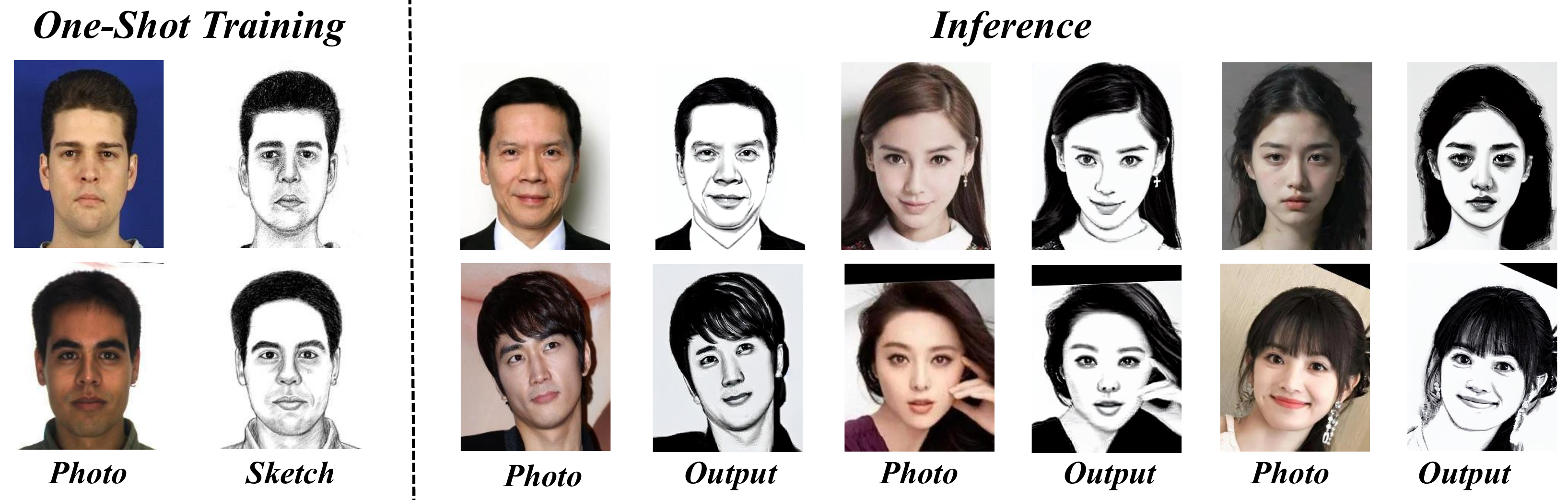}
\caption{An illustration of the proposed in-the-wild face sketch algorithm. In our framework, only one pair of face photo-sketch samples is required for training, and the rest images will be fine for testing.  Confronting with photos featuring different expressions, backgrounds, and sex, our method is capable of handling a diverse range of photos in-the-wild.}
\label{fig4}
\end{figure*}

Existing methods require training on large datasets to generate face sketches, which leads to higher training costs. However, there are only a limited number of artists and face sketches available in most real-world scenarios. 
To address the issue of high training costs, we propose a one-shot face sketch synthesis method based on diffusion models. As shown in Fig.~\ref{fig4}, to model real-world challenges, we only use one pair of face photo-sketch samples for training, and then apply the other samples for inference. \emph{Our method is capable of handling a diverse range of photos, including photos in-the-wild.}

The primary objective of our method is to investigate a text instruction $C _{Text}$ that can accurately describe the editing direction from a face photo to a specific sketch. By using the sketch instruction, our model is enable to generate the expected results with in-the-wild input. Specifically, we set a pair of samples as $\lbrace$$I _{photo}$, $I _{sketch}$$\rbrace$, where $I _{photo}$ is the face photo and $I _{sketch}$ is the sketch. We perform gradient-based optimization on the instruction $C _{Text}$ by leveraging these specific images with rich visual information. To this end, our method is mainly divided into two stages: (1) Optimization through image reconstruction strategies. (2) Optimization through the image editing direction determined. 

\textbf{Face Reconstruction with Diffusion Model.} Usually, diffusion models are trained on a sequence of gradually noisier images over a series of timesteps, e.g., t = 1, . . . , T. With the supervision of the objective function, the diffusion models can acquire the powerful ability to reconstruct images from noise. In this paper, we adopt the image reconstruction strategies of diffusion models to optimize the instruction $C _{Text}$. In this way, $C _{Text}$ is able to encompass the description to recover images from noise, which is a basic and important function of diffusion models for generating images. 

Specifically, we conduct training based on a denoising network. Inspired by Eq.~\ref{eq2}, given a pair of sketch samples $\lbrace$$I _{photo}$, $I _{sketch}$$\rbrace$, we encode and add noise to the sketch $I _{sketch}$. At the same time, we encode the face photo $I _{photo}$  and add it to the denoising network to ensure that the model can follow the pixel-level information of the input photo $I _{photo}$. Then, without fine-tuning the model, we train the noise prediction ability of the model by training the text instruction $C _{Text}$.

Hence, for the sketch sample pair $\lbrace$$I _{photo}$, $I _{sketch}$$\rbrace$, we define the image reconstruction loss $\mathcal{L} _{rec}$ as follows:

\begin{equation}
\label{eq3}
\mathcal{L}_{rec}=\mathbb{E}_{\mathcal{E}(I _{sketch}), c_{Text},z_{p}, \epsilon \sim \mathcal{N}(0,1), t}\left\|\epsilon-\epsilon_{\theta}\left(z_{s
_{t}}, t, C_{Text},z_{p}\right)\right\|_{2}
\end{equation}

where $z _{s _{t}}$ represents the denoised variable of the sketch $I _{sketch}$ after encoding and adding noise at timestep $t$, and $z _{p}$ represents the latent variable after encoding the photo $I _{photo}$ and adding it to the denoising network.

\textbf{One-shot Sketch Editing Direction Acquisition.}
When given an editing instruction, the model can edit the input image according to the editing direction provided by the instruction. However, if $C _{Text}$ is only optimized based on Eq.~\ref{eq3}, it can only learn the prior information inherent to the reconstructed image. It cannot accurately describe the editing direction from a photo to a sketch, which may affect the generation performance of the model.

Nowadays, Contrastive Language-Image Pre-training (CLIP) ~\cite{clip} has become a robust tool for image editing as it can establish a connection between texts and images. As shown in Fig.~\ref{fig3}(a), we calculate the distance between photos and sketches in the high-dimensional space. We use the distance to optimize $C _{Text}$, enabling it to learn the distinction representation between photos and sketches. Thus, $C _{Text}$ becomes to possess a representation of the editing direction from photo to sketch.

Specifically, given a sample pair $\lbrace$$I _{photo}$, $I _{sketch}$$\rbrace$, we calculate the distance between the CLIP embeddings of $I _{photo}$ and $I _{sketch}$, denoted as $\bigtriangleup _{edit}$. Therefore, $\bigtriangleup _{edit}$ can be expressed as:
\begin{equation}
\label{eq4}
\bigtriangleup_{edit}=CLIP_{I}(I _{sketch})-CLIP_{I}(I _{photo})
\end{equation}
where $CLIP_{I}(\cdot)$ represents the image encoder of CLIP.

Then we use $\bigtriangleup _{edit}$ to optimize the text instruction $C _{Text}$. We choose cosine similarity as the loss function of image editing, denoted as $\mathcal{L} _{edit}$. Therefore, $\mathcal{L} _{edit}$ can be expressed as:
\begin{equation}
\label{eq5}
\mathcal{L}_{edit}=cos(\bigtriangleup_{edit},C_{Text})=\frac{\bigtriangleup_{edit}\cdot C_{Text} }{\left | \bigtriangleup_{edit}\left |  \right | C_{Text} \right | } 
\end{equation}
where cos(·) denotes the cosine function.

Finally, we combine the image reconstruction loss $\mathcal{L} _{rec}$ (Eq.~\ref{eq3}) and the image editing loss $\mathcal{L} _{edit}$ (Eq.~\ref{eq4}) to jointly optimize the instruction $C _{Text}$:
\begin{equation}
\label{eq6}
\mathcal{L}=\lambda _{rec}\mathcal{L}_{rec}+\lambda _{edit}\mathcal{L}_{edit}
\end{equation}
where $\lambda _{rec}$ and $\lambda _{edit}$ are hyperparameters.

In this way, after gradient-based optimization, $C _{Text}$ can not only follow the pixel-level details of the input image but also summarize the information of the sketch image while containing the editing direction from the photo to the sketch. The complete algorithm for acquiring One-shot Face Sketch Instruction $C _{Text}$ is shown in Algorithm~\ref{alg1}.

\begin{algorithm}[t!]
\caption{One-shot Face Sketch Instruction Acquisition, given a pre-trained denoising model $\epsilon _{\theta}$, an encoder $\mathcal{E}$ and a CLIP image encoder $CLIP _{I}$($\cdot$).}
\label{alg1}
\begin{algorithmic}

\State \textbf{Input:}  A sketch sample pair $\lbrace$$I _{photo}$, $I _{sketch}$$\rbrace$, encoded as $z _{p}$, $z _{s}$; Optimization steps N; Timesteps T; Initialized text instruction $C _{Text}$; CLIP embedding distance(between $I _{photo}$ and $I _{sketch}$) $\bigtriangleup _{edit}$; Hyperparameters $\lambda _{rec}$ and $\lambda _{edit}$; Learning rate $\gamma$
\State {\bf Output:} Text instruction $C _{Text}$
\State Optimize text instruction $C _{Text}$
\For{$i\gets1$ to N}
    \State Sample $t$ from $U$($0$, $T$); $\epsilon$ from $N$($0$,$1$) 
    
    \State $z _{s _{t}}$ $\gets$ $z _{s}$+$\epsilon$ at timestep $t$
    
    \State $\mathcal{L}$=$\lambda _{rec}\left\|\epsilon-\epsilon_{\theta}\left(z_{s _{t}}, t, c_{Text},z_{p}\right)\right\|_{2}$+$\lambda_{edit}cos(\bigtriangleup_{edit},C_{Text})$
    \State $C _{Text}$ $\gets$ $C _{Text}$ - $\gamma$$\nabla$$\mathcal{L}$

\EndFor
\State \Return $C _{Text}$
\end{algorithmic}
\end{algorithm}

\textbf{Hybrid Instruction Tuning.}
Once the gradient-based optimized $C _{Text}$ is obtained, we can use it as the input textual instruction for the diffusion model during inference. To control the generation direction of the model more precisely, our method allows for additional information to be embedded during inference to guide the process. As shown in Fig.~\ref{fig3}(b), during training, we only optimize a portion of $C _{Text}$. The optimized part is marked as $<ins>$. Then, during inference, we can manually input additional text information to embed into $C _{Text}$, as shown in Fig.~\ref{fig3}(c). The embedded part is marked as $<optional>$. The final hybrid instruction used for inference can be expressed as:
\begin{equation}
\label{eq6}
Hybrid\quad Instruction = <ins> + <optional>
\end{equation}
where $<ins>$ is the optimized part of $C _{Text}$, and $<optional>$ is the additional part embedded to provide more historical context during inference.

Eventually, we  can use the hybrid instruction composed of the learned $C _{Text}$ and the embedded additional information as the text condition for the pre-trained diffusion model. With the allowance of embedding additional information into the instruction, we can control the direction of image generation more flexibly.

\begin{figure*}[!t]
\centering
\includegraphics[width=\textwidth,height=8cm]{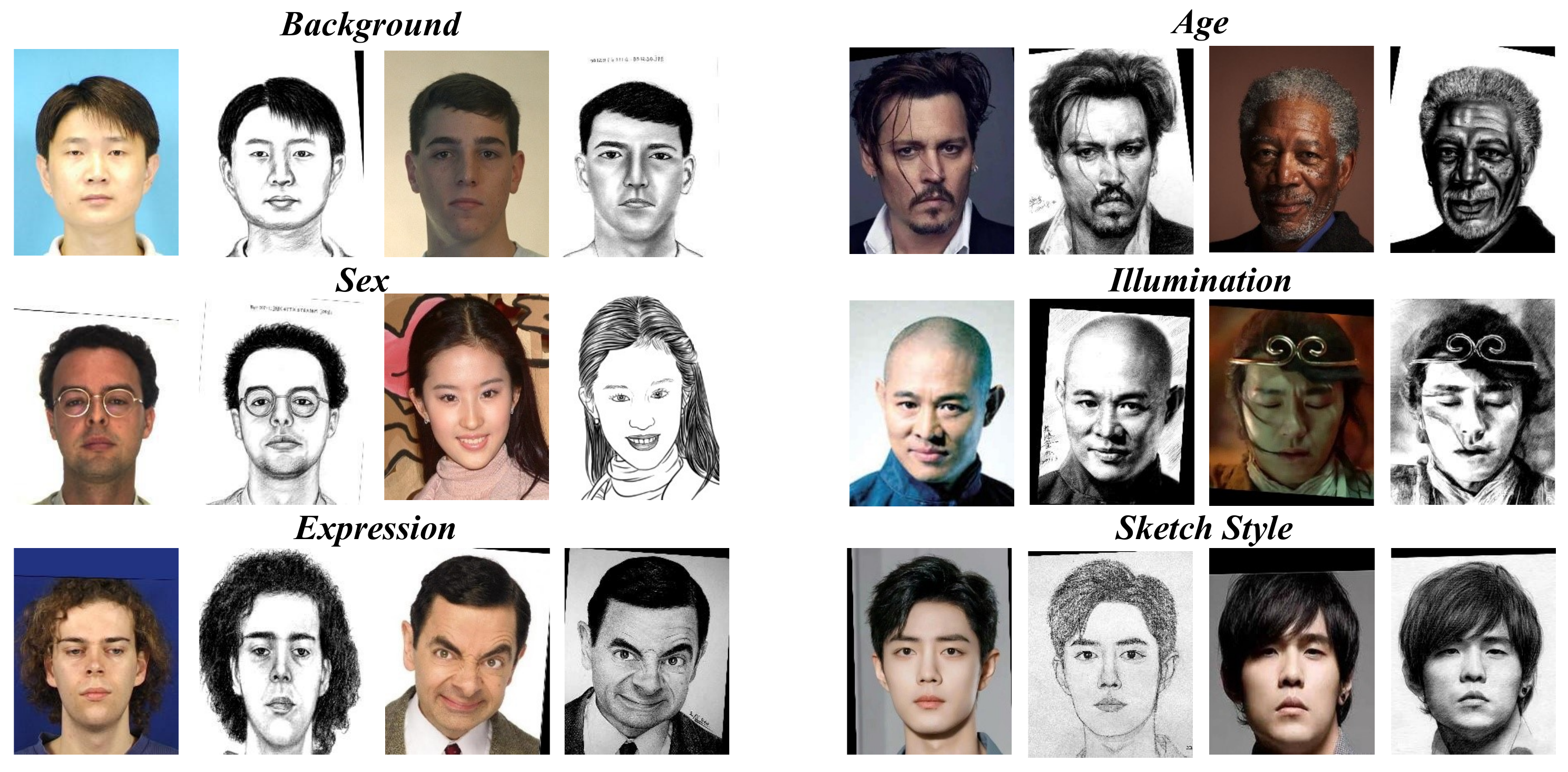}
\caption{An illustration of OS-Sketch dataset. Our dataset contains photos with various backgrounds,  sexes,  expressions,  ages,  illumination as well as sketches in various styles.}
\label{fig6}
\end{figure*}

\section{Experiment}
In this section, we first introduce the implementation details of our method. Then, we describe the datasets and evaluation metrics. Next, we present the experimental results from both qualitative and quantitative perspectives, verifying the effectiveness of the proposed method. Finally, we provide ablation experiments to validate the effectiveness of each module.
\subsection{Implementation Details and Datasets}
In this paper, we use ~\cite{gradient} to automatically generate a text description from the input sketch, serving as the initialized $C _{Text}$. We use a frozen pre-trained model to optimize the instruction $C _{Text}$ at T=1000 timesteps, with the optimization step N set to 12000. We use the AdamW optimizer~\cite{adamw} with learning rate $\gamma$=0.001, $\lambda _{rec}$=4 and $\lambda _{edit}$=0.1. All experiments were conducted on a machine with an NVIDIA 3090 24G GPU.

\textbf{One-shot Face Sketch Dataset.}
\label{OS-Sketch}
To more accurately simulate real-world scenarios and examine the feasibility of sketch synthesis methods in a one-shot context, we introduce a new benchmark called One-shot face sketch datasets (OS-Sketch).  As shown in Fig.~\ref{fig6}, our dataset contains 400 face photo-sketch image pairs, encompassing sketches with various styles and photos with different backgrounds, sexes, expressions, ages, illumination, etc. Our dataset features the most complex range of sketch styles and is the first to be used for one-shot face sketch synthesis, which comprises 100 face photo-sketch image pairs from the simple color backgrounds face sketch dataset~\cite{cufs}, 110 pairs from the dark lighting backgrounds dataset~\cite{cufsf}, 150 pairs from the in-the-wild dataset~\cite{panet} and 40 pairs selected from the internet. The dataset not only contains traditional research photos with a single background but also includes various in-the-wild photos.\emph{We select only ONE pair of images for training, with the REST used for inference.} We assess the effectiveness of the sketch synthesis method by challenging it to process a diverse range of photos and sketches in a one-shot context.  During the inference, the model will encounter numerous independent and identically distributed (i.i.d) and out-of-distribution (OOD) photos. Compared to previous datasets, we believe ours enables more comprehensive evaluation of method performance. All photos and sketches are highly aligned. And all sketches are uniformly converted into high-contrast grayscale images during evaluation. 


\textbf{Evaluation Metrics:} We have selected three evaluation metrics to verify the effectiveness of our method, namely Structural Similarity Index (SSIM)\cite{ssim}, Learned Perceptual Image Patch Similarity (LPIPS)\cite{lpips}, and Fréchet Inception Distance (FID)\cite{fid}.
\subsection{Comparison}
To demonstrate the superiority of our method, we compared it with advanced methods that perform well in face sketch synthesis such as PANet\cite{panet}, DCNP\cite{dcnp}, HIDA\cite{hida} and HiFiSketch\cite{gan9}. Furthermore, to highlight the advantages of the one-shot strategy, we also conduct qualitative and quantitative comparisons with the text-driven diffusion-based style transfer techniques InstructPix2Pix~\cite{insp2p}, PnP-Diffusion\cite{pnp} and pix2pix-zero~\cite{zero} on the OS-Sketch. For a fair comparison, we only used one pair of face photo-sketch samples for training for all methods. In order to control the generation direction of the model more precisely, we select "covert the image color to black and white with a white background" as additional embedded textual information during inference.

\begin{table}[t]
    \centering
    \caption{Quantitative results on the OS-Sketch. The results of the top two performance are highlighted in red and blue, respectively.}
    \label{table1}
    \resizebox{0.5\columnwidth}{!}{%
        \begin{tabular}{lcccc}
            \toprule
            Method & $SSIM\uparrow$ & $LPIPS\downarrow$ & $FID\downarrow$\\
            \midrule
            PANet~\cite{panet} & 0.427 & 0.446 & 116.67  \\
            DCNP~\cite{dcnp} & \textcolor{blue}{\textbf{0.448}} & 0.495 & 111.10  \\
            HIDA~\cite{hida} & 0.408 & \textcolor{blue}{\textbf{0.412}} & 79.13&  \\ 
            HiFiSketch~\cite{gan9} & 0.354 & 0.497 & 128.62  \\
            InstructPix2Pix~\cite{insp2p} & 0.437 & \textcolor{red}{\textbf{0.348}} & \textcolor{blue}{\textbf{78.91}} \\ 
            PnP-Diffusion~\cite{pnp} & 0.388 & 0.545 & 139.59 \\ 
            pix2pix-zero~\cite{zero} & 0.336 & 0.632 & 190.25 \\ 
            Ours & \textcolor{red}{\textbf{0.461}} & \textcolor{red}{\textbf{0.348}} & \textcolor{red}{\textbf{68.46}}&  \\
            \bottomrule
        \end{tabular}%
    }
    \label{table1}
\end{table}

 \begin{figure}[ht]
        \centering
        \includegraphics[width=0.8\textwidth,keepaspectratio]{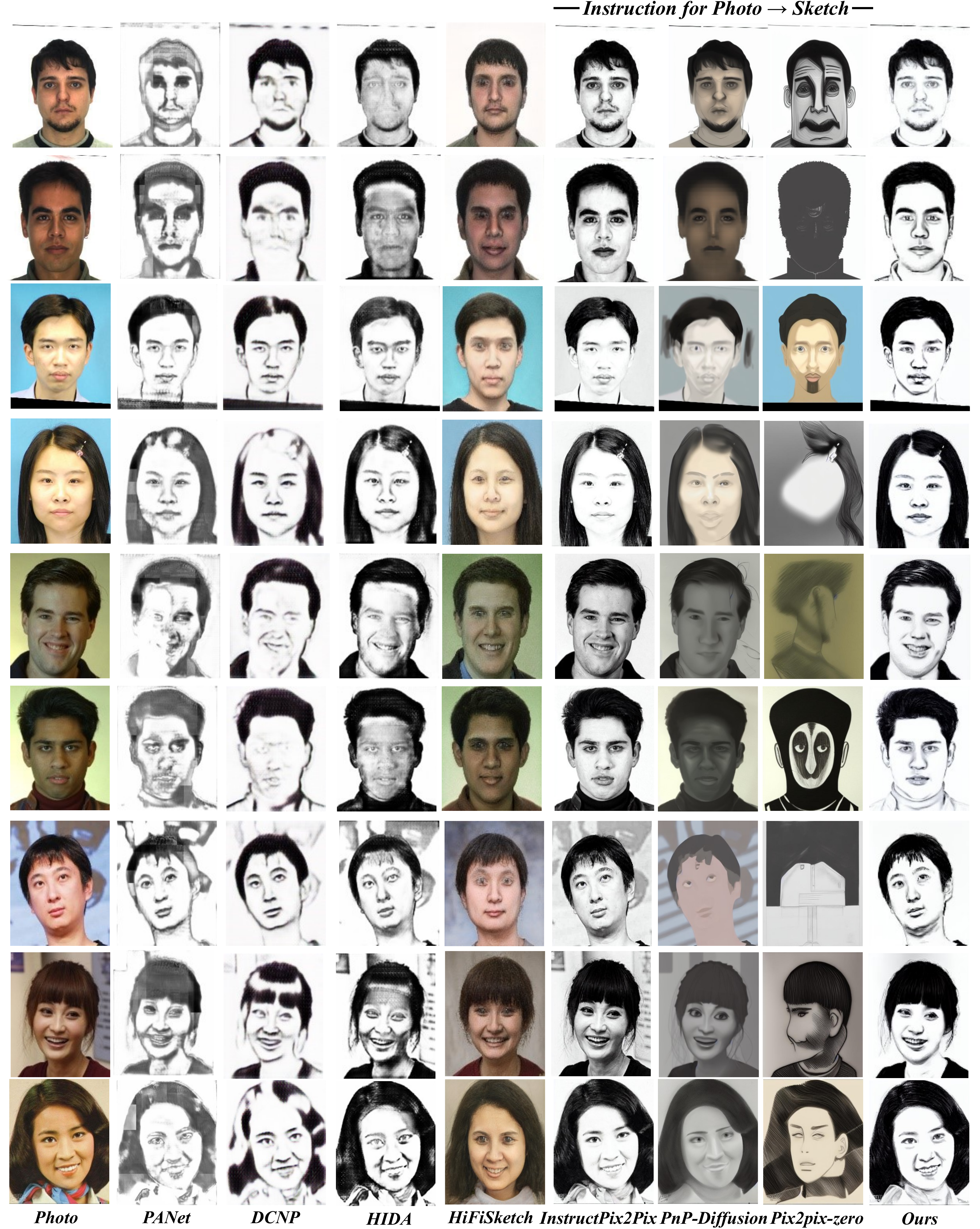}
        \caption{Qualitative comparisons. We conduct a qualitative comparison of our method with other methods on OS-Sketch.  Compared to existing methods, our method has better results.}
        \label{fig7}
        \end{figure}


\textbf{Quantitative and Qualitative Comparisons.} The quantitative results on the OS-Sketch are shown in Table.~\ref{table1}. We achieve the best results. Our method significantly improves upon previous methods in terms of Structural Similarity Index (SSIM)\cite{ssim}, Learned Perceptual Image Patch Similarity\cite{lpips}, and Fréchet Inception
Distance (FID)\cite{fid} metrics. Traditional methods require large datasets for training. In a one-shot context, the models are limited by their training capacity. They can only learn to convert one photo into a sketch with one background, one face, one tone and one style of sketch. When switching to other diverse face photos, there is a substantial decline in the predictive capability of the models. They even fail completely when confronted with out-of-distribution(OOD) photos. As for the text-driven diffusion-based methods, 
they rely on text descriptions solely. However, text descriptions sometimes can be ambiguous, making it difficult to generate precise and appropriate results. Therefore, the results of typical face sketch methods are not satisfactory either. Our method maintains good generation results with one-shot training, \emph{which significantly speeds up the training efficiency, reduces energy costs, and addresses the issue of dataset scarcity.}

The qualitative results of OS-Sketch are shown in Fig.~\ref{fig7}. In a one-shot context, most existing methods fail. The sketches generated by them have issues with structural distortion, colour displacement and blurriness. For the text-driven diffusion-based methods, the generation results are also unsatisfactory due to the difficulty in describing specific sketch generation tasks with solely text descriptions. The generated sketches have issues with facial structure, expressions, and regional colours, which are difficult to control precisely through words. Compared to existing methods, our method has better results. In a one-shot scene, our method can generate complete and realistic sketches from face photos with different backgrounds, sexes, illumination, etc. Our approach can convert both independent and identically distributed (i.i.d) and out-of-distribution (OOD) photos
into sketches at the same time. It remains effective even when applied to photos with complex backgrounds. In contrast, traditional methods sometimes even fail to generate a complete face. The experiments demonstrate the superiority of our method, which achieves impressive results with only one pair of samples for training, significantly improving experimental efficiency.

\begin{figure*}[t]
\centering
\includegraphics[width=\linewidth]{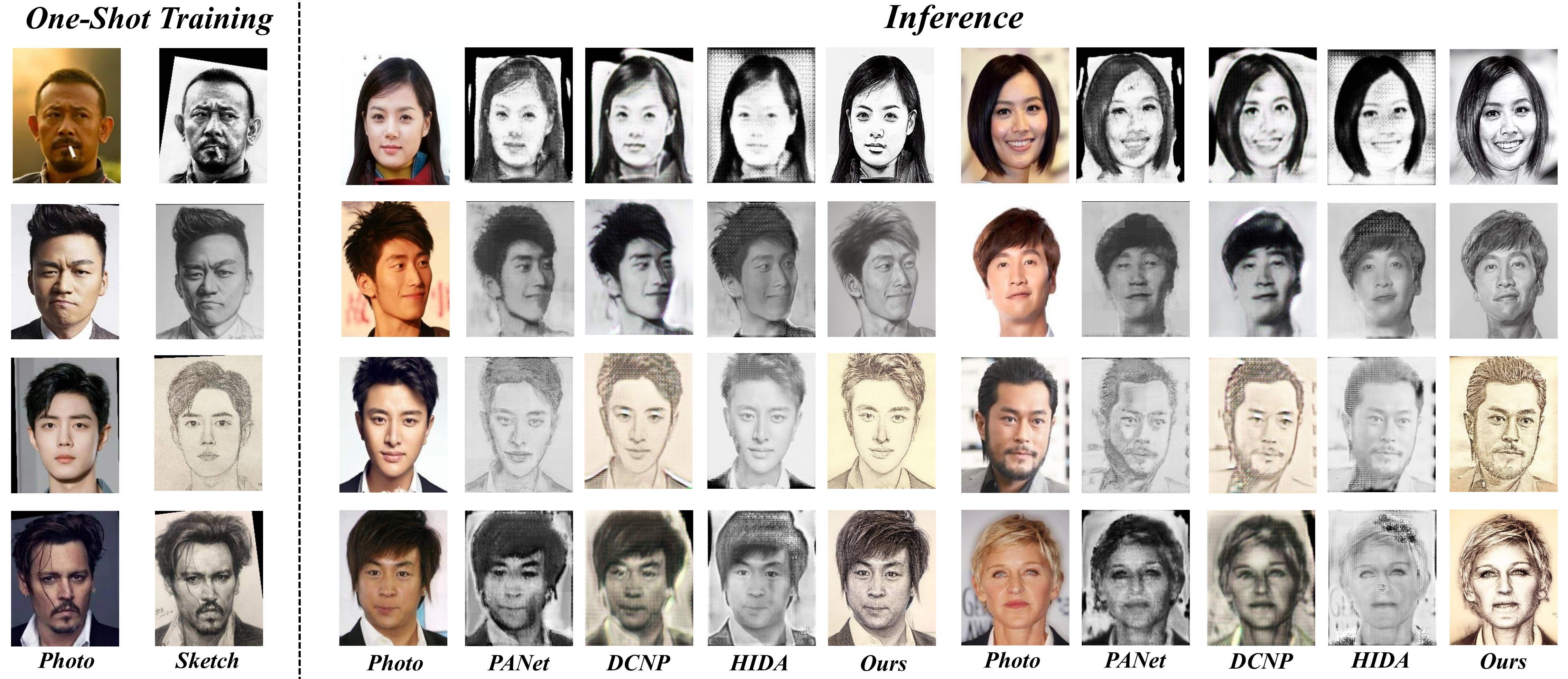}
\caption{We exhibit an in-the-wild experiment by using 4 challenging samples selected from the internet for out-of-distribution sketch synthesis. Experiments demonstrate that our method achieves good performance for sketches in-the-wild. It not only indicates that we can directly conduct research on sketches in-the-wild instead of relying on existing datasets, but also means that we can manipulate the style of sketches more flexibly. }
\label{fig9}
\end{figure*}
\textbf{Efficiencies of Training and Inference Phrases.} Moreover, we conduct an efficiency analysis of our method compared to other approaches. Specifically, we compared the training efficiency of our method with the state-of-the-art face sketch synthesis techniques and compared the inference efficiency with the diffusion-based methods. Both the training and inference comparisons are conducted using a single image. As shown in Table.~\ref{table2}, compared to PANet~\cite{panet}, DCNP~\cite{dcnp}, HIDA~\cite{hida} and HiFiSketch~\cite{gan9}, our method not only achieves top-tier training 
speed but also delivers superior generation results in terms of LPIPS~\cite{lpips} and FID~\cite{fid}. As shown in Table.~\ref{table3}, compared to PnP-Diffusion~\cite{pnp} and pix2pix-zero~\cite{zero}, although they do not require any training, our inference speed surpasses theirs. And our method generates results with greater precision. The experimental results confirm that our method offers higher accuracy and practical usability.

\begin{table}[t]
    \caption{Comparison of \emph{training} efficiency. The
best results are shown in bold and suboptimal results are underlined.}
    \label{table2}
    \centering
    \small
    \resizebox{0.7\columnwidth}{!}
    {
    \begin{tabular}{|c|c|c|c|c|c|}
    \hline
   Method                   & PANet~\cite{panet} & DCNP~\cite{dcnp} & HIDA~\cite{hida}  & HiFiSketch~\cite{gan9} &  Ours                \\ \hline \hline
    Training times (S)          & \textbf{6062.20}   & 11954.76  & 10361.46  & 71058.99 &\underline{7616.35}
    \\ \hline
    $LPIPS\downarrow$           &0.446   &0.495  &\underline{0.412} &0.497  &\textbf{0.348}   
    \\ \hline
    $FID\downarrow$              &116.67   &111.10 &\underline{79.13} &128.62  &\textbf{68.46}
    \\ \hline 
    \end{tabular}
    }
    
\end{table}

\begin{table}[t]
    \caption{Comparison of \emph{inference} efficiency. The
best results are shown in bold.}
    \label{table3}
    \centering
    \small
    \resizebox{0.7\columnwidth}{!}
    {
    \begin{tabular}{|c|c|c|c|c|}
    \hline
 Method               & PnP-Diffusion~\cite{pnp}  & pix2pix-zero~\cite{zero}  &  Ours    \\ \hline \hline
    Inference times (S)           & 194.21   & 24.93   & \textbf{5.99}               
    \\ \hline
     $LPIPS\downarrow$           &0.545   &0.632    &\textbf{0.348}   
    \\ \hline
    $FID\downarrow$              &139.59   &190.25   &\textbf{68.46}
     \\ \hline                            

    \end{tabular}
    }
    
\end{table}

\textbf{Out-of-Distribution Comparison.} OS-Sketch contains 40 photos selected from the internet and corresponding sketches drawn by bloggers or amateurs. To further explore our method, we select 4 original sketches with distinctive characteristics for training and testing the rest images. The qualitative results are shown in Fig.~\ref{fig9}.

As can be seen from the visualization results, our method is capable of producing sketches with reasonable quality even with face photo-sketch pairs selected from the internet. Our method can generate realistic sketch textures and maintains a tone similar to the training samples, whereas other methods fail in a one-shot context. Experiments demonstrate the superiority of our method. In a one-shot context, our method can still adapt to out-of-distribution (OOD) styles of sketches. This attempt also suggests that in future research, specialized datasets are not required any more. Instead, we can directly select random face photo-sketch pairs from the internet. This finding greatly reduces the dependence on datasets and allows us to control the style of the sketches more flexibly. 

\subsection{Ablation Study}

\begin{figure}[t]
\includegraphics[width=0.5\textwidth]{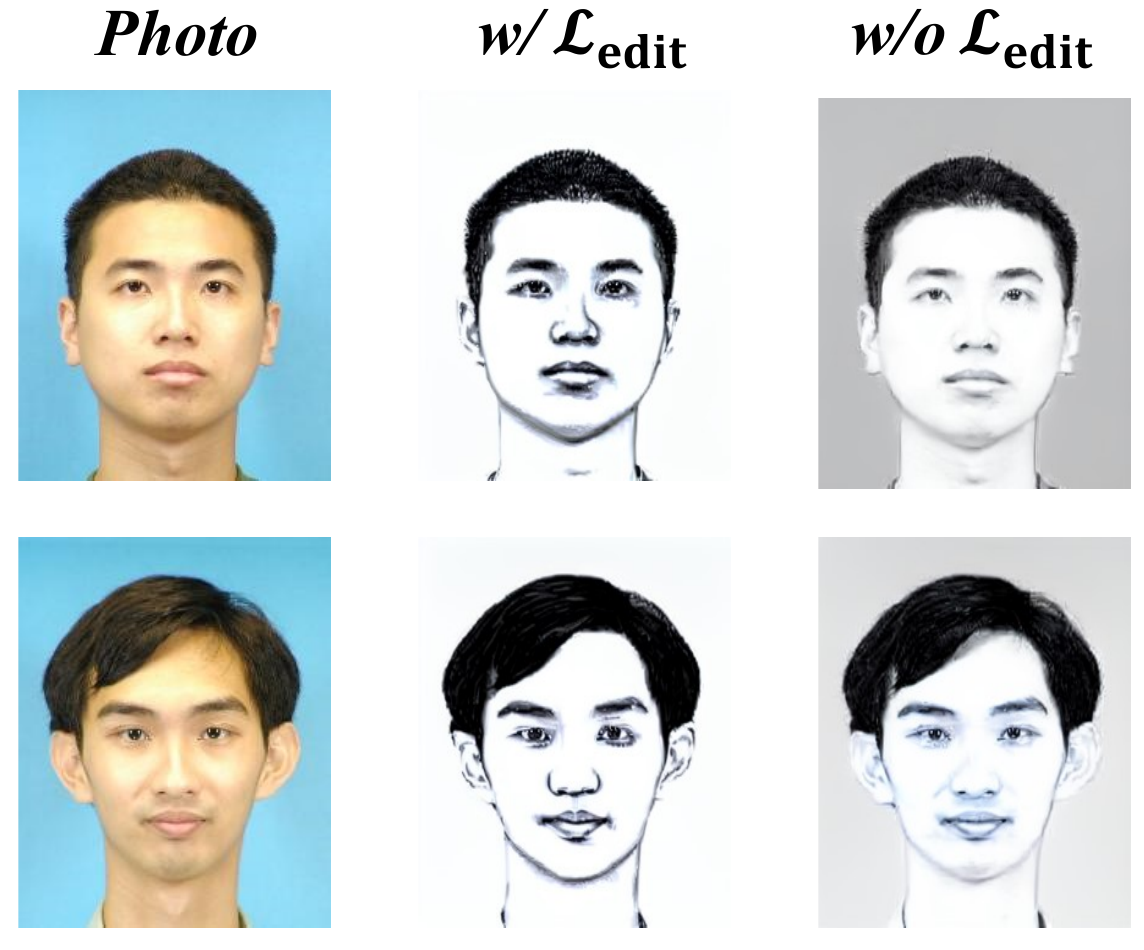}
\caption[something short]{Comparison of the generation results of  "w/ $\mathcal{L} _{edit}$" and "w/o $\mathcal{L} _{edit}$". Without the use of $\mathcal{L} _{edit}$, the generated image has not been truly converted into a sketch image. Experiments demonstrate the effects of sketch prompt acquisition.}
\label{fig10}
\vspace{-3mm}
\end{figure}

\textbf{Effects of Generative Diffusion Prior.} As shown in Fig.~\ref{fig10}, we demonstrate the difference in effect between with and without $\mathcal{L} _{edit}$ . Without the use of $\mathcal{L} _{edit}$, the learned $C _{Text}$ lacks information on the editing direction from a photo to a sketch. The model only learns the information of the sketch image itself and does not understand the expected editing direction. Therefore, the generated image has not been truly converted into a sketch. Thus, it is essential to incorporate $\mathcal{L} _{edit}$  to obtain a more satisfactory editing direction.

\begin{figure}[t]
\includegraphics[width=0.7\textwidth]{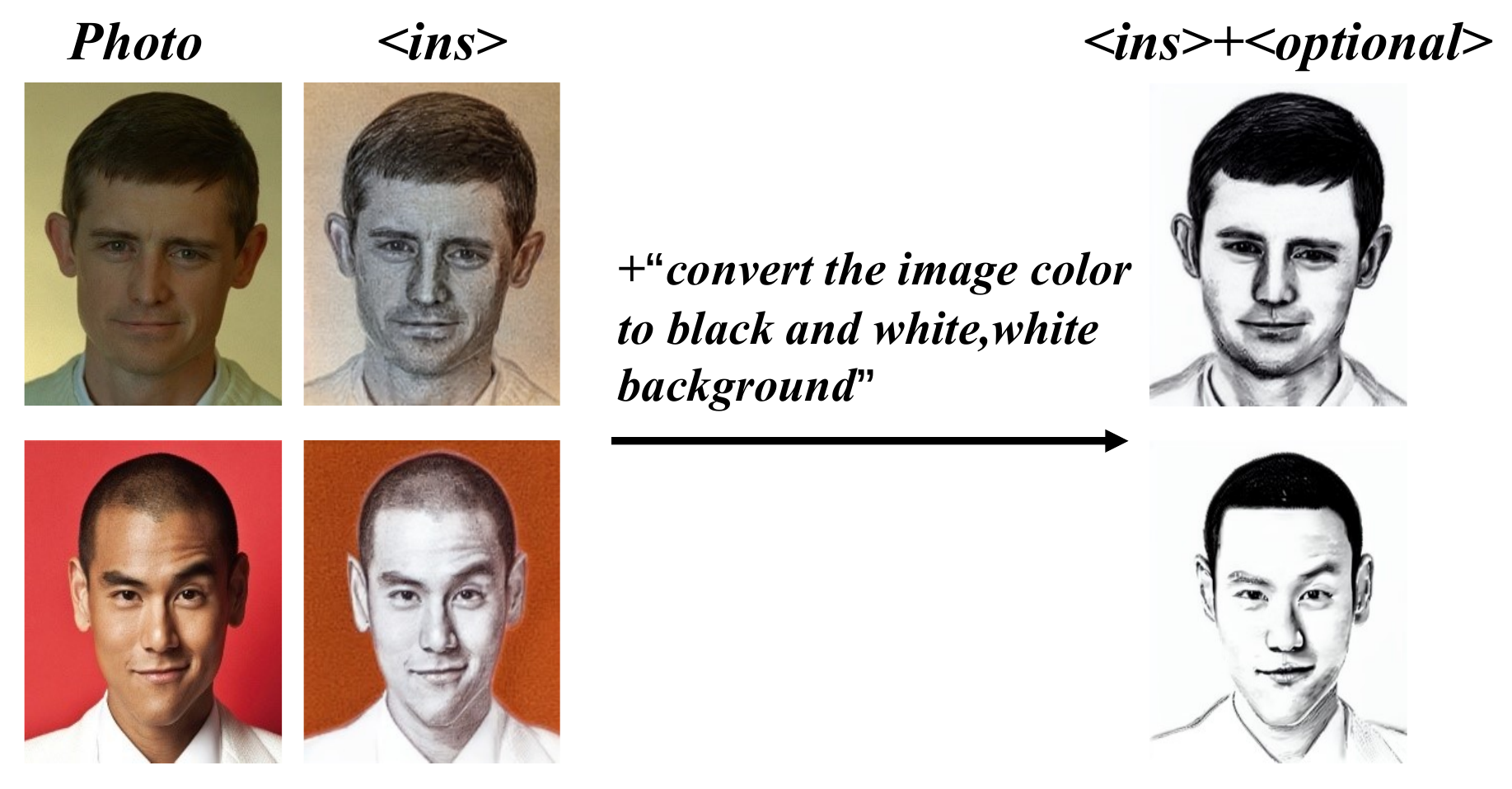}
\caption{Comparison of the generation results of guidance with hybrid instruction and with solely $<ins>$ . Experiments demonstrate that instruction with embedded additional information have a more accurate result.}
\label{fig11}
\vspace{-3mm}
\end{figure}

\textbf{Effects of Hybrid Instruction Tuning.} As shown in Fig.~\ref{fig11}, we demonstrate the difference in effect between guidance with hybrid instruction and with solely $<ins>$ without additional information input. It is clear that without manually inputting extra information, the tone of the images generated by the model sometimes may have some deviations. When selecting "convert the image color to black and white, white background" as the additional embedded textual information (i.e., <optional>), the model generates the desired results. The experiments confirm that with the guidance of hybrid instruction tuning, the model can better control the generation effect and accurately convert a face photo into the desired sketch.

\section{Limitation and Conclusion}
In this paper, we introduce a new framework for face sketch synthesis called One-shot Face Sketch Synthesis. To more accurately simulate real-world scenarios and evaluate the feasibility of our method in a one-shot context, we introduce a new benchmark called OS-Sketch. It includes sketches of various styles and photos with different backgrounds, expressions, sexes, illumination, etc. Extensive experiments demonstrate the excellent performance of our framework in both quantitative and qualitative results. Our method can generate realistic sketches and achieve face sketch synthesis in-the-wild using only one pair of samples for training. In one-shot scenes, our method maintains high consistency between the sketches and the original photos. Additionally, our method enables the model to generate sketches in corresponding styles by training on sketches selected from the internet. Our method is highly suitable for face sketch synthesis, significantly reducing training costs and improving training efficiency.

However, our method is based on the text-to-image diffusion model, which still requires corresponding prompt and instructions.  Without additional textual information embeddings, sometimes we may not have the expected results. Besides, there may be limitations in handling all types of sketch styles or achieve more precise sketch textures due to the operational mode of text-to-image diffusion models. In the future, more advanced solutions should be proposed to further enhance efficiency, optimize textures or other details.

\bibliographystyle{ACM-Reference-Format}
\bibliography{ref}










\end{document}